\documentclass[oneside,letterpaper,10pt]{article}
\pdfoutput=1

\usepackage[letterpaper,total={6.5in, 9in},top=1in]{geometry}

\usepackage[plain]{fancyref}
\usepackage[dvipsnames]{xcolor}
\usepackage{graphicx}
\usepackage{hyperref}
\hypersetup{
    colorlinks=true,
    linkcolor=blue,
    filecolor=magenta,      
    urlcolor=cyan,
}
\usepackage{xspace}
\usepackage{booktabs}
\usepackage{flushend}

\title{BenchPress: Analyzing Android App Vulnerability Benchmark Suites}

\author{Joydeep Mitra \hspace{1cm} Venkatesh-Prasad Ranganath \hspace{1cm} Aditya Narkar\\
Kansas State University, USA\\
\{joydeep,rvprasad,avnarkar\}@ksu.edu
}

\date{Created: January 29, 2019.  Revised: September 19, 2019.}

\begin{document}

\maketitle

\begin{abstract}
    In recent years, various benchmark suites have been developed to evaluate the efficacy of Android security analysis tools. The choice of such benchmark suites used in tool evaluations is often based on the availability and popularity of suites and not on their characteristics and relevance. One of the reasons for such choices is the lack of information about the characteristics and relevance of benchmarks suites.

    In this context, we empirically evaluated four Android specific benchmark suites: DroidBench, Ghera, ICCBench, and UBCBench. For each benchmark suite, we identified the APIs used by the suite that were discussed on Stack Overflow in the context of Android app development and measured the usage of these APIs in a sample of 227K real world apps (coverage). We also compared each pair of benchmark suites to identify the differences between them in terms of API usage. Finally, we identified security-related APIs used in real-world apps but not in any of the above benchmark suites to assess the opportunities to extend benchmark suites (gaps).  

    The findings in this paper can help 1) Android security analysis tool developers choose benchmark suites that are best suited to evaluate their tools (informed by coverage and pairwise comparison) and 2) Android app vulnerability benchmark creators develop and extend benchmark suites (informed by gaps).
\end{abstract}

\newcommand{\ie}{i.e.,\xspace}
\newcommand{\eg}{e.g.,\xspace}
\newcommand{\etal}{et al.\xspace}
\newcommand{\citeneed}[1]{(cite #1)}
\newcommand{\jm}[1]{\textcolor{red}{[#1]}}
\newcommand{\rv}[1]{\textcolor{blue}{[#1]}}
\newcommand{\fillme}{\textcolor{red}\textbf{FILL ME}}
\newcommand*{\addheight}[2][0.5ex]{%
  \raisebox{0pt}[\dimexpr\height+(#1)\relax]{#2}%
}
\newcommand{\droidbench}{DroidBench\xspace}
\newcommand{\ghera}{Ghera\xspace}
\newcommand{\iccbench}{ICCBench\xspace}
\newcommand{\ubcbench}{UBCBench\xspace}

\section{Introduction}

\subsection{Motivation}
Effectiveness of Android security analysis tools is evaluated with benchmarks and real-world apps.  The effectiveness of static taint analysis tools like AmanDroid \cite{Wei:2014}, FlowDroid \cite{Arzt:2014}, HornDroid \cite{Calzavara:2017}, and IccTA \cite{Li:2015} has been evaluated by applying them to benchmarks from \droidbench, \iccbench, and \ubcbench \cite{Qiu:2018} benchmark suites and comparing tool verdicts with benchmark labels that indicate the presence/absence of specific vulnerability or malicious behavior.

Such tool evaluations have used benchmarks without evaluating the authenticity and the representativeness of the benchmarks.   \emph{Authenticity} is the truthfulness of the claim about the presence/absence of a vulnerability or malicious behavior in a benchmark (Section 2.2.2 in \cite{Mitra:PROMISE17}).  \emph{Representativeness} is the similarity between the manifestation/occurrence of a vulnerability in a benchmark and in real-world apps (Section 3 in \cite{Ranganath:EMSE19}).  Consequently, the usefulness of findings from these evaluations is diminished in terms of the ability of tools and techniques to detect vulnerabilities or malicious behaviors (due to authenticity) and the general applicability of tools and techniques (due to representativeness).

Recently, there have been two efforts focused on the authenticity of benchmarks.  Mitra and Ranganath \cite{Mitra:PROMISE17} created \ghera, a suite of demonstrably authentic Android app vulnerability benchmarks, to address the issue of authenticity.
They also established the representativeness of \ghera benchmarks (in terms of API usage) \cite{Ranganath:EMSE19}.
Pauck \etal \cite{Pauck:2018} developed ReproDroid, a tool to help verify the authenticity of Android app vulnerability benchmarks.  They found that not all claims about the presence/absence of vulnerabilities in benchmarks in DIALDroid, \droidbench, and \iccbench benchmark suites were true.

It is common in other communities to study and characterize benchmarks. In the program analysis community, Blackburn \etal \cite{Blackburn:2006} developed and used metrics based on static and dynamic properties of programs to characterize and compare the DaCapo benchmarks with SPEC Java benchmarks \cite{SPECJava:URL}.  Isen \etal \cite{Isen:2008} measured several properties of embedded Java benchmarks and how well they represent real-world mobile apps. In the systems community, Pallister \etal \cite{Pallister:2013} characterized benchmarks based on the energy consumption properties of embedded platforms.  In the database community, such assessments have been around since 1990s \cite{Gray:Book93}. However, such close scrutiny of benchmark suites have not occurred in the Android security community.

Motivated by the aforementioned efforts to analyze and characterize benchmarks, we undertook an effort to assess the representativeness of multiple Android app vulnerability benchmark suites.\ie how well does a benchmark suite represent real-world apps?

\subsection{Research Questions}

The objective of our effort is to answer the following research questions:
\begin{itemize}
    \item \textbf{RQ1} \textit{In general, do Android app vulnerability benchmark suites use APIs that are used by real-world apps and discussed by Android app developers?} The question is aimed at understanding the representativeness of benchmark suites and the relevance of the APIs used to capture vulnerabilities in benchmark suites.  The answer to this question can help associate an element of confidence to benchmarks and, consequently, to tool evaluations that use such benchmarks.


     \item \textbf{RQ2} \textit{In the context of security, do Android app vulnerability benchmark suites use APIs that are used by real-world apps and discussed by Android app developers?}  Similar to RQ1, this question is intended to understand the representativeness and relevance of the security-related APIs used by benchmark suites but in the specific context of security. 

    \item \textbf{RQ3} \textit{How do the considered benchmark suites differ in terms of API usage?} The purpose of this question is to identify the common and unique APIs between benchmark suite pairs. The answer to this question can help tool developers choose appropriate benchmark suites to test/evaluate their tools.

    \item \textbf{RQ4} \textit{Do real-world apps use security-related APIs not used by any benchmark suite?}  The purpose of this question is to identify gaps between existing benchmark suites and the real-world apps in terms of security-related APIs.  The answer to this question can steer security analysis efforts towards unexplored Android APIs to possibly uncover new vulnerabilities and enhance existing benchmark suites.

\end{itemize}




\subsection{Contributions}
In this paper, we make the following contributions:

\begin{itemize}
    \item Provide empirical evidence about the representativeness of four Android app vulnerability benchmark suites.
    \item Identify gaps between the evaluated benchmark suites and real world apps in terms of APIs that are used in real world apps but not in the benchmark suites.
    \item Extend and improve the framework for empirical evaluation of Android app vulnerability benchmarks introduced by Ranganath \& Mitra \cite{Ranganath:EMSE19} because we believe this framework can be used by other researchers to conduct similar studies in other domains as well.
\end{itemize}

In addition to these contributions, we hope this effort will spark the interest of the empirical software engineering community to study Android app vulnerability benchmarks and help improve Android app security.

The remainder of the paper is structured as follows. Section \ref{sec:approach} outlines the metric of representativeness along with the benchmark suites and the real world app sample used in the study.  Section \ref{sec:experiment} describes the experiment to measure representativeness. 
Sections \ref{sec:rq1}-\ref{sec:rq4} discuss the answers to posed research questions. Section \ref{sec:threats} describes the threats to the validity of the experiment. Section \ref{sec:relwork} describes related work. Section \ref{sec:artifacts} provides information about the artefacts used in this effort. Section \ref{sec:summary} summarizes the findings from this effort.


\section{Concepts and Subjects}
\label{sec:approach}

\subsection{API usage as a measure of representativeness}

Representative vulnerability benchmarks should have two
aspects. First, they should capture vulnerabilities that occur in
the real world. Second, the manifestation of vulnerabilities in
representative benchmarks should be similar (if not identical)
to that in real-world apps.

Ranganath and Mitra \cite{Ranganath:EMSE19} observed this challenge while establishing the representativeness of \ghera benchmarks.  So, they introduced the notion of using API usage as a weak but general measure of representativeness of benchmarks.  They reasoned ``the likelihood of a vulnerability occurring in real-world apps is directly proportional to the number of real-world apps using the Android APIs involved in the vulnerability''.  Consequently, to measure the representativeness of benchmarks, they measured how often APIs used in benchmarks were used in real-world apps.

In this evaluation, we use the above notion and a similar approach to measure the representativeness of benchmarks.

\subsection{Benchmarks}

For this study, we considered 4 benchmark suites related to Android app vulnerabilities: \textit{DroidBench, Ghera, IccBench,} and \textit{UBCBench}.

\textit{DroidBench} \cite{Pauck:2018} contains 211 benchmarks. Each benchmark is an Android app that captures zero or more information leak vulnerabilities. The vulnerabilities captured in DroidBench primarily stem from Inter-Component Communication (ICC) feature of Android and general features of Java.

\textit{Ghera} \cite{Mitra:PROMISE17} contains 60 benchmarks that capture mostly known Android app vulnerabilities along with few unknown Android app vulnerabilities. Each benchmark includes 3 Android apps: 1) a \emph{benign} app that contains vulnerability \emph{x}, 2) a \emph{malicious} app that exploits vulnerability \emph{x} in the benign app, and 3) a \emph{secure} app that does not contain vulnerability \emph{x} and thus cannot be exploited by the malicious app. In this evaluation, we considered only the benign apps from each benchmark and we will refer to them as Ghera benchmarks in the rest of this paper. Unlike in DroidBench, the vulnerabilities in Ghera stem from different Android features including ICC.

\textit{IccBench} \cite{Pauck:2018} contains 24 benchmarks.  Each benchmark is an Android app that captures zero or more information leak vulnerabilities. IccBench focuses on capturing vulnerabilities that stem from communcation between apps via ICC.

\textit{UBCBench} \cite{Qiu:2018} contains 16 benchmarks. Each benchmark is an Android app that captures at most one information leak vulnerability. UBCBench captures information flow vulnerabilities primarily stemming from ICC and SharedPreferences\footnote{A SharedPreference is a file that stores key-value pairs and can be private to an app or shared} features of Android and general features of Java.

\subsection{Real World Apps}

We collected 700K apps from AndroZoo \cite{Allix:MSR16} in March 2019. From this set of 700K apps, we curated a set of 473K apps that target API levels 19 thru 27.  An API level uniquely identifies the framework API revision offered by a version of the Android platform.  In an Android app, the \emph{minimum API level} is the least framework API version required by the app and \emph{target API level} is the framework API version targeted by the app.  For this evaluation, we initially picked target API level 19 thru 27 because most benchmarks targeted these API levels.  However, we later discovered that Android currently does not support API levels 19 thru 22.  Therefore, to make the evaluation current, from the set of 473K apps, we retained only apps that target API levels 23 thru 27.  Hence, we ended up with a sample of 226K real-world Android apps.  \Fref{tab:real_world_app_sample} provides the distribution of this sample across considered target API Levels.

\begin{table}
    \centering
    \ifdef{\TopCaption}{
      \caption{Distribution of target API levels in the sample of real world apps}
    }{}
    \begin{tabular}{cc}
        \hline
         Target API level & \# Real World Apps\\
         \hline
         23 & 146K\\
         24 & 18K\\
         25 & 16K\\
         26 & 29K\\
         27 & 17K\\
         \hline
         Total & 226K\\
         \hline
    \end{tabular}
    \ifundef{\TopCaption}{
      \caption{Distribution of target API levels in the sample of real world apps}
    }{}
    \label{tab:real_world_app_sample}
\end{table}

\section{Experiment}
\label{sec:experiment}

\subsection{Preparing the benchmarks}
\label{subsec:prep-bnchmrks}

By design, each Android app is bundled as a self-contained APK file that contains all code and resources necessary to execute the app but are not provided by the underlying Android framework.  However, \textit{due to the build process of Android apps, the APKs may contain unnecessary code and resources (e.g. unused methods)}.  So, ProGuard \cite{ProGuard:URL} tool can be used as part of the Android app build process to remove unnecessary artifacts from APKs.

Every benchmark suite considered in this evaluation provides pre-built APKs and source files for each of its benchmarks.  The pre-built APKs provided by \droidbench, \iccbench, and \ubcbench contain unnecessary code and resources.  Also, the benchmarks do not have the same minimum and target API levels.  Specifically, \droidbench benchmarks have minimum API level 8 and target API level 14 thru 24, \ghera benchmarks have minimum API level 22 and target API level 27, \iccbench benchmarks have minimum and target API level 25, and \ubcbench benchmarks have minimum and target API level 19.

Since we wanted to measure the representativeness of benchmark suites and compare them based on API usage, we needed to control for the effects of unnecessary APIs and API level on the findings of the evaluation. Therefore, we rebuilt every benchmark from its source with minimum API level set to 23, target API level set to 27, using appcompat support library version 27.1.1, and using Proguard to remove unnecessary APIs. We chose API levels 23 thru 27 because they are currently supported by Android.

We ensured the rebuilt benchmarks were indeed supported by API levels 23 thru 27 by executing each benchmark on an emulator running Android 23 and 27.  As part of the execution, we manually interacted with the app until no further interaction was possible.  Often, this meant interacting with various widgets on a screen and navigating to various screens in an app.

If the benchmark or app crashed, then we recorded the crash. \Fref{tab:prep_benchmarks_info} lists the total number of benchmarks in each suite, the number of benchmarks that we were able to successfully build, and the number of benchmarks that crashed during execution.  In this evaluation, we considered all benchmarks that could be built successfully including the ones that crashed.  We considered the crashed benchmarks because we were unsure if they crashed due to a vulnerability intentionally captured in the benchmark or other reasons such as change of API levels.

\subsubsection*{Observations}

From \Fref{tab:prep_benchmarks_info}, we see that, most benchmarks not only build but also execute on the currently supported versions of Android (even when they were not designed to run on those versions) -- out of 311 benchmarks across all benchmark suites, only 35 crashed during execution and only 10 could not be built successfully.
So, \textit{while most benchmarks were not explicitly designed to run on recent versions of Android, they are well supported by recent versions of Android}.

The 10 benchmarks (from \droidbench) that failed to build imply \textit{some benchmarks are not supported by recent versions of Android}; an important aspect that should be considered when using \droidbench to evaluate effectiveness of Android security analysis tools.

For the 35 benchmarks (32 from \droidbench and 3 from \ubcbench) that crashed when executed on emulators running Android 23 and 27, we re-executed pre-built counterparts of these benchmarks on an emulator running the version of Android originally targeted by the benchmarks.  Interestingly, all of the benchmarks crashed during re-execution.  \textit{This raises the question ``are these 35 benchmarks from \droidbench and \ubcbench valid and, hence, authentic?''}


\begin{table}
    \centering
    \ifdef{\TopCaption}{
      \caption{Total No. of benchmarks in each benchmark suite along with the No. of benchmarks that built successfully with minimum API level 23 and target API level 27 and crashed on an emulator running Android 23 and 27.}
    }{}
    \begin{tabular}{lrrr}
        \hline
         Benchmark & \# Total & \# Benchmarks & \# Benchmarks\\
         Suite & benchmarks & built successfully & crashed\\
         \hline
         \droidbench & 211 & 201 & 32\\
         \ghera & 60 & 60 & 0\\
         \iccbench & 24 & 24 & 0\\
         \ubcbench & 16 & 16 & 3\\
         \hline
    \end{tabular}
    \ifundef{\TopCaption}{
      \caption{Total No. of benchmarks in each benchmark suite along with the No. of benchmarks that built successfully with minimum API level 23 and target API level 27 and crashed on an emulator running Android 23 and 27.}
    }{}
    \label{tab:prep_benchmarks_info}
\end{table}

\subsection{API-based App Profiling}
\label{sec:profiling}
The APK file of an app contains an XML-based manifest file and a DEX file that contains the code and data (\ie resources, assets) of the app. Android apps use various features of the underlying Android framework via XML-based manifest files and published Android programming APIs.  We refer to these published Android programming APIs and the XML elements and attributes of manifest files collectively as APIs.

We use an adaptation of the method used by Ranganath and Mitra \cite{Ranganath:EMSE19} to determine the APIs used by the sample of real-world Android apps and by the benchmarks.

In this method, for each app, the elements and attributes in its manifest along with all callback methods and all methods that were used but not defined in the app are considered while ignoring obfuscated methods, \ie methods with single character names.  Further, to make apps comparable, for overridden methods and fields, class hierarchy analysis is used to consider the overridden methods and fields as opposed to the overriding methods and fields.
Of these APIs, only APIs whose fully qualified name (FQN) contained the prefix \textit{android, com.android, java,} or \textit{org} are considered because the method focuses on measuring representativeness in terms of Android APIs.

Numerous APIs are commonly used in almost all Android apps and are related to aspects (\eg UI renderin
g) that are not the focus of vulnerability benchmarks related to Android apps. To avoid their influence
 on the experiment, such APIs are ignored while determining the APIs used in the benchmarks. For this p
urpose, we created a baseline app with minimum and target API levels as 23 and 27, respectively.  This
app did not exhibit any interesting functionality but contained graphical widgets commonly used by Andr
oid apps. Out of the 1847 APIs used in this baseline app, we manually identified 1586 APIs as commonly
used in Android apps; almost all of them were basic Java APIs or related to UI rendering and XML proces
sing. For each benchmark suite, we ignored these 1586 APIs from the list of recorded APIs (from the pre
vious step) to create the set of \textit{considered} APIs. Even in the \textit{considered} set, we identified APIs orthogonal to the focus of these benchmarks, \eg \emph{android.graphics}. So, we filtered out these APIs to create a set of \textit{filtered} APIs. This additional filtering did not change the API sets drastically as can be seen from the 2nd and 3rd columns in \Fref{tab:api_usage_info}.

\begin{table}
  \centering
  \ifdef{\TopCaption}{
    \caption{Number of APIs used by the benchmark suites}
  }{}
  \begin{tabular}{lrrrrr}
    \hline
    Repo & \multicolumn{5}{c}{Number of APIs}\\
    \cmidrule(r){2-6}
    & Total & Considered & Filtered & Relevant & Security\\
    \hline
    \droidbench & 2188 & 837 & 798 & 769 & 744\\
    \ghera & 1906 & 565 & 518 & 504 & 494\\
    \iccbench & 185 & 102 & 70 & 70 & 70\\
    \ubcbench & 751 & 127 & 99 & 98 & 96\\
    \hline
  \end{tabular}
  \ifundef{\TopCaption}{
    \caption{Number of APIs used by the benchmark suites}
  }{}
  \label{tab:api_usage_info}
\end{table}

\subsection{Using Android app developer discussions in Stack Overflow to identify relevant and security-related APIs}
\label{sec:SoF}


Ranganath and Mitra \cite{Mitra:PROMISE17} considered the set of filtered APIs as \emph{relevant} to Android app development.  From this set, they manually identified a subset of (\emph{security-related}) APIs as related to Android security.  Since the experimenter's subjectivity and bias could influence the findings of the evaluation via this manual identification, we decided to use Stack Overflow data to identify security-related APIs.  Further, we also decided to use Stack Overflow data to prune the set of filtered APIs into the set of relevant APIs.

For this purpose, we used a snapshot of Stack Overflow posts from March 2019 \cite{SoFArchive:URL} as follows:

\paragraph{Identified Android related posts} We considered all posts from the Stack Overflow snapshot with the \textit{Android} tag. This resulted in 1.2 million posts.\footnote{Once we considered every Stack Overflow posts with the \textit{Android} tag, we wondered if we were missing out on posts that were related to Android but did not have the \textit{Android} tag.  To account for such posts, we collected all tags (including synonyms) from \textit{Android Stack Exchange}, a forum for discussing issues related to Android.  From this set of 1347 tags, we removed 433 tags (including synonyms) that had been considered in the 1.2 million Stack Overflow posts.  Of the remaining 914 tags, we removed tags related to companies as they were not related to Android app development.  Of the remaining 445 tags, only 6 tags -- instapaper, pebble, sdhc, task-management, xfat -- were associated with \textit{Stack Overflow} posts.  Upon examination, we decided that none of the six tags were related to Android app development and hence we ignored them.  Since this extra step did not influence the number of posts related to Android, we did not use it in our experiment.}

\paragraph{Identified Android security related posts} There is no readily available information in Stack Overflow data to identify security related posts.  So, we used the data from \textit{Security Stack Exchange}, a forum for discussing software security issues, to identify security related posts on Stack Overflow \cite{SoFArchive:URL}. We gathered all tags from Security Stack Exchange as the set of \textit{security-related tags}.  We also added the \emph{security} tag to this set.  From the Stack Overflow posts that had \textit{Android} tag, we identified posts with at least one \textit{security-related} tag.  We ended up with a set of 460K posts related to \emph{Android security}.

\paragraph{Filtered posts based on API levels}  Since API level 23 was released in 2015, we considered only posts that had some activity -- created, were answered, edited, commented on, voted on, or marked as favorite or accepted -- on or after 2015.  An API that did not garner interest after 2015 is likely deprecated in API levels 23 thru 27 or well understood by developers.  In either case, we deemed such APIs as irrelevant to our evaluation.  After this filtering, we ended up with 831K Android related posts and 318K Android security related posts.

\paragraph{Identified relevant and security-related APIs} If the class name and the method/field name of an API co-occurred in a post, then we considered the post as discussing the API.  Based on this notion, if a filtered API was discussed in an Android related post, then the API was deemed as a  \textit{relevant API}, \ie relevant to Android app development.  Similarly, if a filtered API was discussed in an Android security related post, then we deemed it as a \textit{security-related API}.

\subsection{Measuring Representativeness with API Usage Percentage}

For each benchmark suite, for each corresponding relevant API, we calculated the percentage of real world apps that used the API. We did the same for each security-related API.

\subsection{Examining Gap between Real World Apps and Benchmarks}
\label{sec:gapExp}
Of the 2118K APIs used by the apps in our real world app sample, we ignored the APIs used in all benchmarks in \droidbench, \ghera, \iccbench, and \ubcbench. Of the remaining 2115K APIs, we ignored APIs related to UI and third party libraries because such APIs are not the focus of the benchmarks being evaluated. The remaining 26K APIs were spread across 31 unique package-prefixes\footnote{A package-prefix of an API is a substring of its package name from the first character to the character before the second occurring separator. For example, the package-prefix of the package \textit{android/app/Activity} is \textit{android/app} or the package-prefix of the package \textit{android/telephone/gsm/SMSManager} is \textit{android/telephony}.}. From this set, we identified security-related APIs using Stack Overflow posts with at least one security-related tag (as described in \Fref{sec:SoF}).  Based on the frequency of Stack Overflow posts discussing an API, we selected the top 10 APIs across 27 package-prefixes and manually examined them to see if they could lead to a vulnerability and, consequently, to the creation of a new benchmark.  We ignored 4 package-prefixes -- \textit{android/print, android/printservices, android/inputmethodservice,} and \textit{android/Manifest} retrospectively since we realized that they contain APIs pertaining to features (\eg UI and printing services) that are not the focus of the considered benchmark suites.

\section{Answering RQ1}
\label{sec:rq1}

\textit{In general, do Android app vulnerability benchmark suites use APIs that are used by real-world apps and discussed by Android app developers?}

\noindent
\begin{figure*}[t]
    \centering
\begin{tabular}{cc}
      \addheight{\includegraphics[width=8cm]{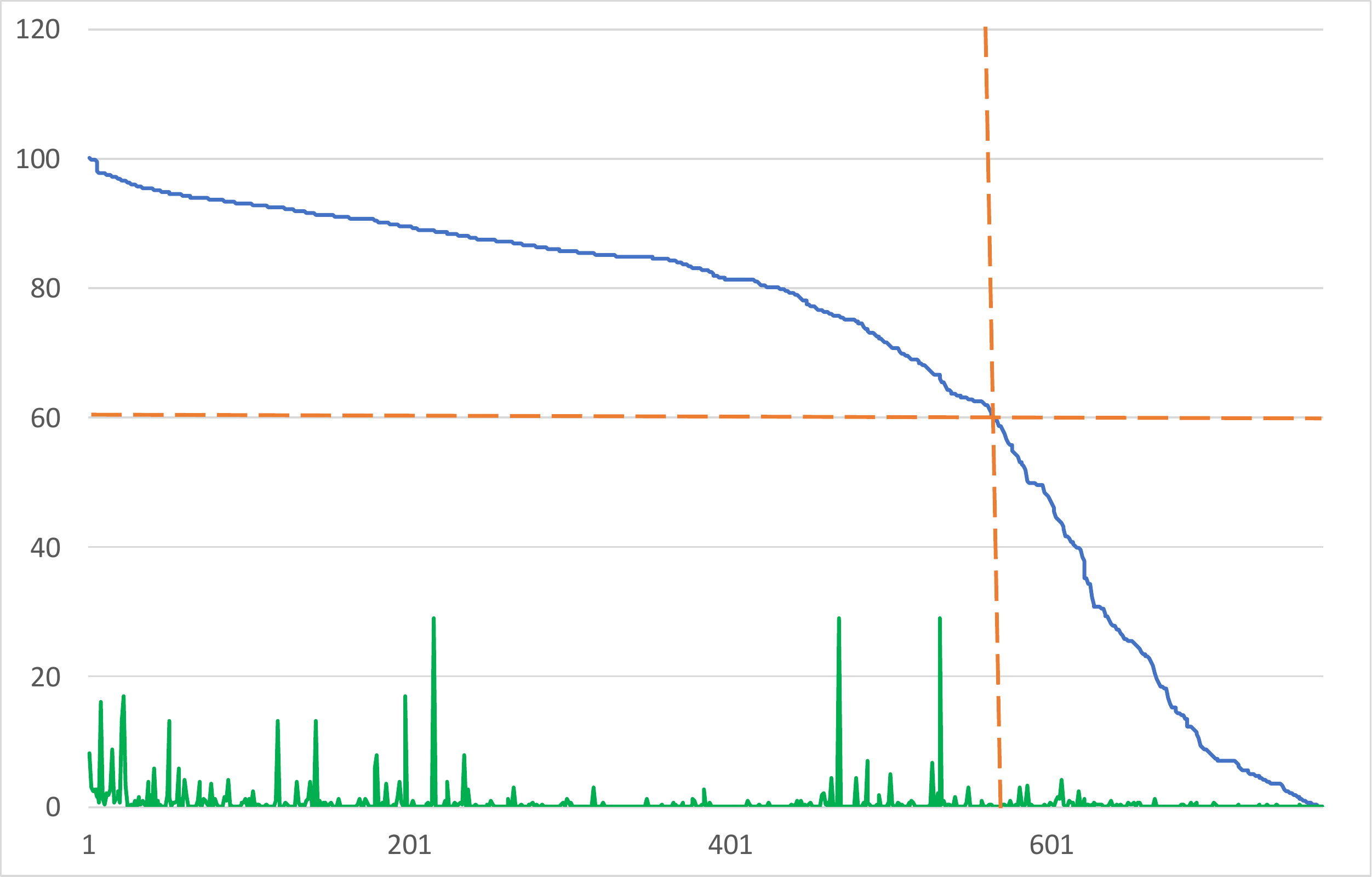}} &
      \addheight{\includegraphics[width=8cm]{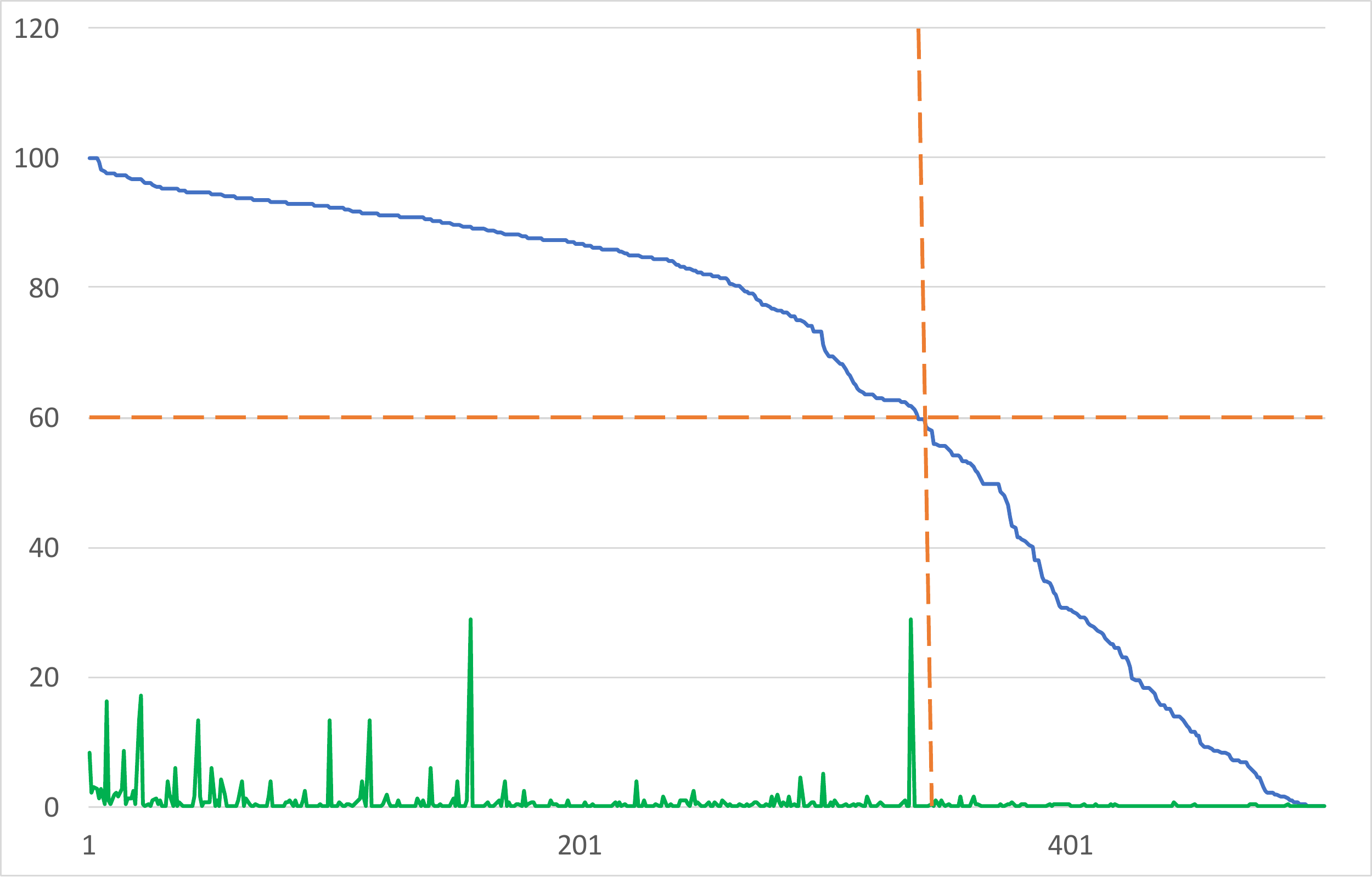}} \\
      \droidbench & \ghera \\
      \addheight{\includegraphics[width=8cm]{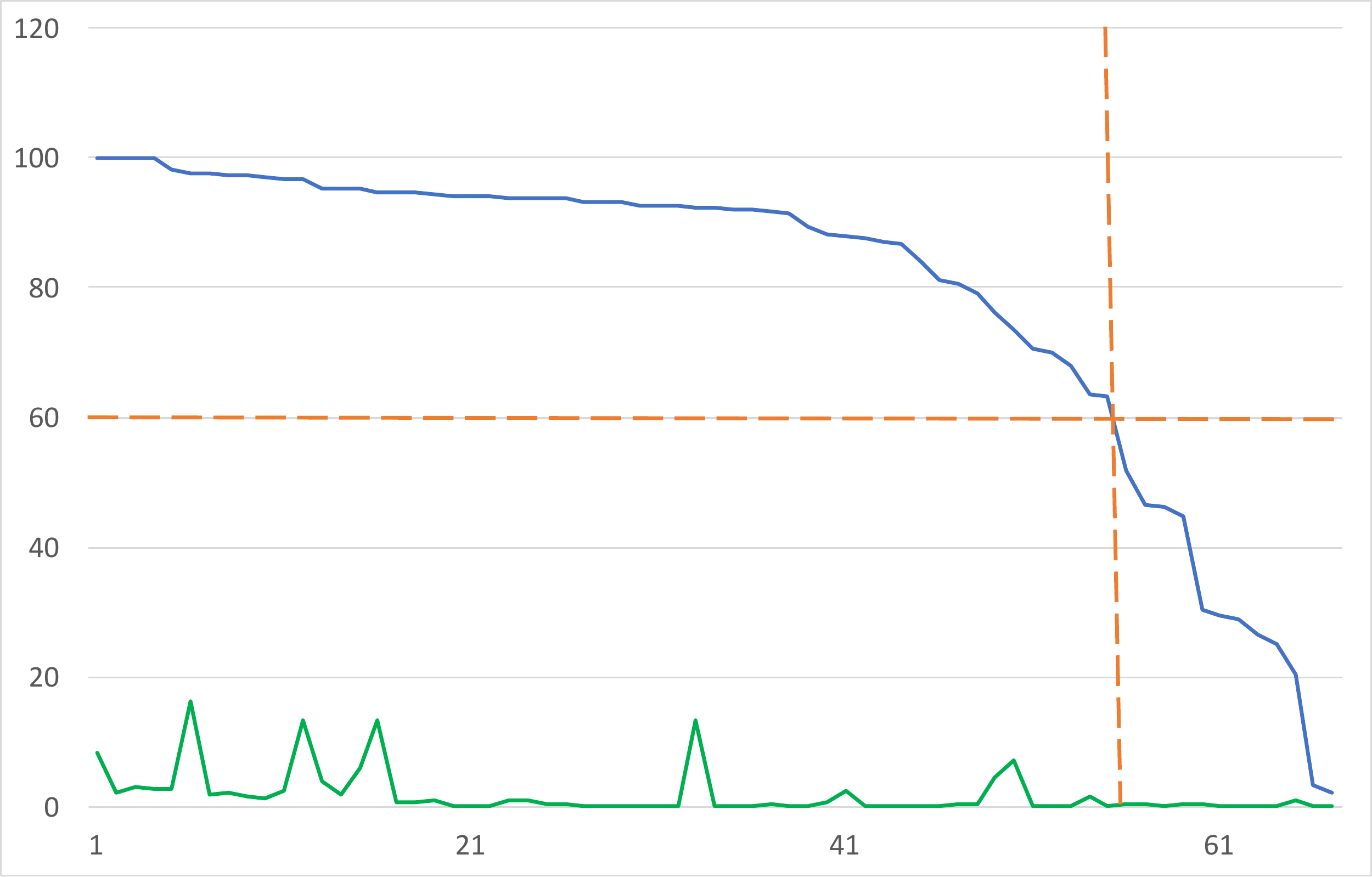}} &
      \addheight{\includegraphics[width=8cm]{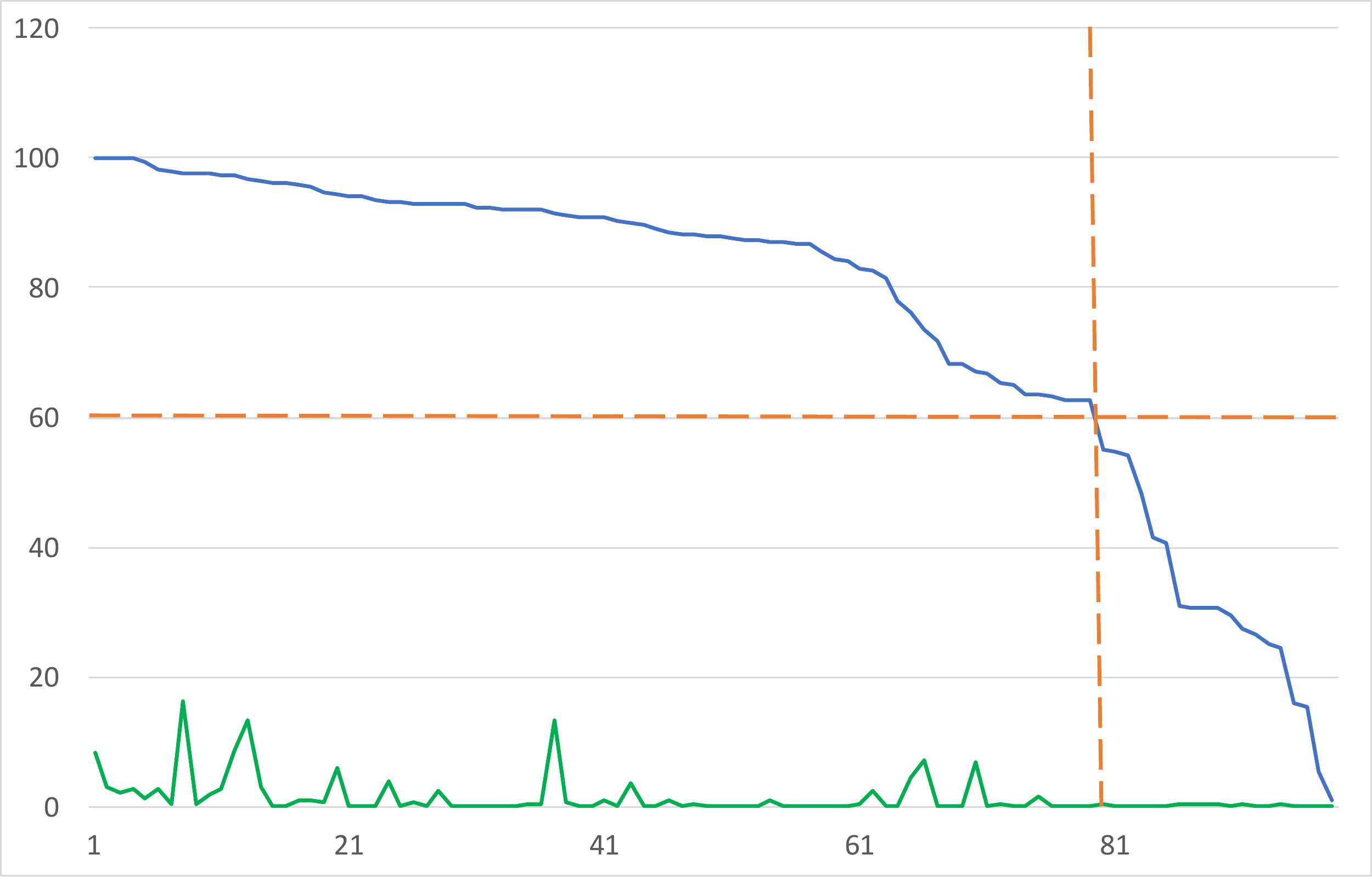}} \\
      \iccbench & \ubcbench \\
      \multicolumn{2}{c}{Relevant APIs in decreasing order of use percentage for API Levels 23-27}\\
      \multicolumn{2}{c}{\textcolor{blue}{\textbf{-----} }\% Real-World apps using a relevant API}\\
      \multicolumn{2}{c}{\textcolor{ForestGreen}{\textbf{-----} }\% Android related Stack Overflow Posts discussing a relevant API}\\
\end{tabular}
    \caption{Percentage of real-world apps that use a relevant API and the percentage of Stack Overflow posts that discuss a relevant API in a benchmark suite.}
\label{fig:repr-graphs}
\end{figure*}

For each benchmark suite, the corresponding graph in \Fref{fig:repr-graphs} shows the percentage of real-world apps using a relevant API that is used by the suite along with the percentage of Stack Overflow posts discussing the same API.

\Fref{tab:5num-summ-posts} and \Fref{tab:5num-summ-sec-posts} list the five-number summary of the percentage of posts discussing relevant and security-related APIs, respectively.

\paragraph*{\droidbench} We found that only 29 of the 798 filtered APIs used by \droidbench are not discussed by any Android related Stack Overflow post. The remaining 769 APIs are discussed in at least 1 post.  As seen in \Fref{tab:5num-summ-posts}, more than half the 769 relevant APIs used by \droidbench are discussed by at least 385 posts.  Therefore, \textit{APIs used by \droidbench are being discussed by Android app developers}.

The graph for \droidbench in \Fref{fig:repr-graphs} shows that all relevant APIs are used by real-world apps and 562 (73\%) relevant APIs are used by more than 60\% of real-world apps. Therefore, \textit{\droidbench is representative of real-world apps in terms of API usage}.

\paragraph*{\ghera} We found that only 14 of the 518 filtered APIs used by \ghera are not discussed in any Android related Stack Overflow post. The remaining 504 (relevant) APIs are discussed in at least one \textit{Stack Overflow} post. As seen in \Fref{tab:5num-summ-posts}, more than half the 504 relevant APIs are discussed by at least 845 posts. Therefore, \textit{APIs used by \ghera are being discussed by Android app developers}.

The graph for \ghera in \Fref{fig:repr-graphs} shows that all relevant APIs are used by real-world apps and 340 (67\%) relevant APIs are used by more than 60\% of real-world apps. Therefore, \textit{\ghera is representative of real-world apps in terms of API usage}.


\paragraph*{\iccbench} We found that all of the 70 filtered APIs used by \iccbench are discussed by at least one Android related Stack Overflow post. As seen in \Fref{tab:5num-summ-posts}, more than half of the 70 relevant APIs used by \iccbench are discussed by at least 2446 posts. Therefore, \textit{APIs used by \iccbench are being discussed by Android app developers}.

The graph for \iccbench in \Fref{fig:repr-graphs} shows that all relevant APIs are used by real-world apps and 55 (78\%) relevant APIs are used by more than 60\% of real-world apps. Therefore, \textit{\iccbench is representative of real-world apps in terms of API usage}.

\paragraph*{\ubcbench} We found that 98 of the 99 filtered APIs used by \iccbench are discussed by at least one Android related Stack Overflow post. As seen in \Fref{tab:5num-summ-posts}, more than half the 98 relevant APIs used by \ubcbench are discussed by at least 1406. Therefore, \textit{APIs used by \ubcbench are being discussed by Android app developers}.

The graph for \ubcbench in \Fref{fig:repr-graphs} shows that all relevant APIs are used by real-world apps and 79 (81\%) relevant APIs are used by more than 60\% of real-world apps. Therefore, \textit{\ubcbench is representative of real-world apps in terms of API usage}.

In short, \emph{\droidbench, \ghera, \iccbench, and \ubcbench are all representative of real-world apps in terms of API usage.}

\subsection{Discussion}

Comparatively, \droidbench (769) and \ghera (504) use more than five times the number of relevant APIs used by \ubcbench (98) and \iccbench (70).  In terms of the percentage of relevant APIs used by more than 60\% of the real-world apps, \droidbench (562) and \ghera (340) use more than four times the number of relevant APIs used by \ubcbench (79) and \iccbench (55).  So, in terms of coverage of APIs used by real-world apps, \droidbench and \ghera fare better than \ubcbench and \iccbench.

In \Fref{fig:repr-graphs}, most of the spikes in the line corresponding to Android related Stack Overflow posts are associated with relevant APIs that are used by more than 60\% of the real-world apps. Hence, the benchmarks are not only representative but also relevant since they are using APIs that are not only used by a large number of real-world apps but are also being discussed widely by Android app developers.

\section{Answering RQ2}
\label{sec:rq2}

\textit{In the context of security, do Android app vulnerability benchmark suites use APIs that are used by real-world apps and discussed by Android app developers?}

As shown in \Fref{tab:api_usage_info}, the number of \textit{security-related} APIs as deemed by Stack Overflow data is similar to the number of relevant APIs. For example, in \iccbench, the number of security-related APIs is identical to the number of relevant APIs.  Moreover, as seen in \Fref{tab:5num-summ-posts}, the distribution of posts discussing security-related APIs is similar to that of relevant APIs as can be seen from the five-number summary. Consequently, the observations for relevant APIs carries over to security-related APIs.

This observation differs from that made by Ranganath \& Mitra \cite{Ranganath:EMSE19} because they \textit{manually} identified security-related APIs while we used Stack Overflow data to identify security-related APIs. They deemed 601 APIs in \ghera as relevant and identified that 117 of the them were related to security. On the contrary, we deemed 504 APIs as relevant and discovered that 494 of them were related to security. Interestingly, all the 117 security-related APIs they identified, were deemed as security-related by us as well.

\paragraph*{Caveat} The numbers in \Fref{tab:5num-summ-posts} suggest that almost all of the relevant APIs used by a benchmark suite are related to security.  These numbers are based on our approach of identifying relevant and security-related APIs using Stack Overflow data as explained in \Fref{sec:SoF}. 
While our approach considers the occurrence of APIs in posts to identify an API as relevant or security-related, it does not consider the context in which the API occurs in posts, \ie an API can occur in a post as part of a code snippet that is being discussed but yet not be discussed in the post.  Hence, our approach can conservatively identify APIs that are irrelevant or not related to security as relevant or security-related.  Therefore, the number of relevant APIs and security-related APIs used by a benchmark suite is likely lower than the numbers reported here.

\begin{table*}
  \centering
  \ifdef{\TopCaption}{
    \caption{Five-Number summary of 831K relevant posts discussing APIs in a benchmark suite}
  }{}
  \begin{tabular}{lrrrrrrrrrr}
    \hline
    Repo & \multicolumn{10}{c}{\% (No.) of relevant posts discussing APIs}\\
    \cmidrule(r){2-11}
    & \multicolumn{2}{c}{Min} & \multicolumn{2}{c}{Q1} & \multicolumn{2}{c}{Median} & \multicolumn{2}{c}{Q3} & \multicolumn{2}{c}{Max}\\
    \cmidrule(r){2-3} \cmidrule(r){4-5} \cmidrule(r){6-7} \cmidrule(r){8-9} \cmidrule(r){10-11}
    \droidbench & 0.0001 & (1) & 0.0068 & (57) & 0.04 & (385) & 0.29 & (2423) & 29.0 & (240K)\\
    \ghera & 0.0001 & (1) & 0.0156 & (130) & 0.10 & (845) & 0.44 & (3668) & 29.0 & (240K)\\
    \iccbench & 0.0007 & (6) & 0.0588 & (489) & 0.29 & (2446) & 1.71 & (14242) & 15.3 & (127K)\\
    UBCBench & 0.0001 & (1) & 0.0348 & (289) & 0.17 & (1406) & 0.95 & (7880) & 15.3 & (127K)\\
    \hline
  \end{tabular}
  \ifundef{\TopCaption}{
    \caption{Five-Number summary of 831K relevant posts discussing APIs in a benchmark suite}
  }{}
  \label{tab:5num-summ-posts}
\end{table*}

\begin{table*}
  \centering
  \ifdef{\TopCaption}{
    \caption{Five-Number summary of 318K security-related posts discussing APIs in a benchmark suite}
  }{}
  \begin{tabular}{lrrrrrrrrrr}
    \hline
    Repo & \multicolumn{10}{c}{\% (No.) of security-related posts discussing APIs}\\
    \cmidrule(r){2-11}
    & \multicolumn{2}{c}{Min} & \multicolumn{2}{c}{Q1} & \multicolumn{2}{c}{Median} & \multicolumn{2}{c}{Q3} &\multicolumn{2}{c} {Max}\\
    \cmidrule(r){2-3} \cmidrule(r){4-5} \cmidrule(r){6-7} \cmidrule(r){8-9} \cmidrule(r){10-11}
    \droidbench & 0.0003 & (1) & 0.007 & (22) & 0.05 & (168) & 0.31 & (997) & 35.0 & (111K)\\
    \ghera & 0.0003 & (1) & 0.020 & (65) & 0.11 & (359) & 0.49 & (1559) & 35.0 & (111K)\\
    \iccbench & 0.0009 & (3) & 0.056 & (!79) & 0.273 & (869) & 1.55 & (4940) & 14.3 & (45K)\\
    UBCBench & 0.0006 & (2) & 0.033 & (106) & 0.20 & (640) & 1.08 & (3458) & 14.3 & (45K)\\
    \hline
  \end{tabular}
  \ifundef{\TopCaption}{
    \caption{Five-Number summary of 318K security-related posts discussing APIs in a benchmark suite}
  }{}
  \label{tab:5num-summ-sec-posts}
\end{table*}

\section{Answering RQ3}
\label{sec:rq3}

\textit{How do the considered benchmark suites differ in terms of API usage?}

We answer RQ3 by presenting observations based on pairwise comparison of the benchmark suites.

For each pairwise comparison, \Fref{tab:api_compare_relevant_info} shows the number of filtered APIs common and unique to the compared benchmarks.

\begin{table}
  \centering
  \ifdef{\TopCaption}{
    \caption{Filtered APIs based pairwise comparison of benchmark suites}
  }{}
  \begin{tabular}{lrrr}
    \hline
    Benchmark Suite & Common & APIs unique & APIs unique\\
    Pair (X-Y) & APIs & to X & to Y\\
    \hline
    \droidbench-\ghera & 344 & 454 & 174\\
    \droidbench-\iccbench & 67 & 731 & 3\\
    \droidbench-\ubcbench & 89 & 709 & 10\\
    \ghera-\iccbench & 42 & 476 & 28\\
    \ghera-\ubcbench & 85 & 433 & 14\\
    \iccbench-\ubcbench & 16 & 54 & 83\\
    \hline
  \end{tabular}
  \ifundef{\TopCaption}{
    \caption{Filtered APIs based pairwise comparison of benchmark suites}
  }{}
  \label{tab:api_compare_relevant_info}
\end{table}


\subsection{\droidbench vs \ghera}
\paragraph*{Observation 1} \droidbench uses 1.5 times the number of APIs used by \ghera but it contains almost 3 times the number of benchmarks in \ghera; see \Fref{tab:api_usage_info}. Therefore, \textit{the difference in API usage between \droidbench and \ghera is not comparable to the difference in the number of benchmarks in them}. This is most likely because \droidbench focuses on heavily ICC (depth) whereas \ghera focuses on ICC and other Android features (breadth). Consequently, the benchmarks in \droidbench use more common APIs compared to the benchmarks in \ghera.

\paragraph*{Observtion 2} \droidbench uses 454 APIs not used by \ghera; see \Fref{tab:api_compare_relevant_info}.  438 of these APIs were identified as relevant using Stack Overflow data.  Moreover, 299 (68\%) relevant APIs are used by at least 60\% real-world apps. Of the 438 relevant APIs, approximately 100 are not specific to Android but related to Java. Of the remaining APIs, close to 50\% are related to ICC. \textit{Since a large number of APIs unique to \droidbench are related to ICC, evaluations of Android app vulnerability detection tools, especially the ones that focus on ICC, should consider \droidbench}.

Moreover, 344 APIs are common to \droidbench and \ghera. Of these, 331 APIs are relevant and 280 (85\%) APIs are used by at least 60\% of real-world apps. 200 of the 344 APIs are related to ICC.  This is not surprising as \droidbench focuses on ICC. Therefore, tools focusing on detecting vulnerabilities stemming from ICC can use either \droidbench or \ghera for evaluation. However, \textit{since only 65 of the 174 APIs unique to \ghera are related to ICC, such tools should prefer \droidbench over \ghera}.


\paragraph*{Observation 3} From \Fref{tab:api_compare_relevant_info}, we see \ghera uses 174 APIs that are not used by \droidbench. 173 of these APIs were identified as relevant using Stack Overflow data. Of these relevant APIs, 69 (40\%) APIs are used by at least 60\% real-world apps.  Further, 93 (54\%) of the 173 relevant APIs are related to Android features such as web, crypto, storage, and networking features. \textit{Since \ghera benchmarks capture vulnerabilities stemming from the use of APIs not related to ICC, evaluations of tools that detect vulnerabilities not related to ICC should consider \ghera.}

\paragraph*{Observation 5} All benchmarks in \ghera capture vulnerabilities that can be reproduced and exploited on API Levels 23 thru 27. However, \droidbench benchmarks were designed to run on older API Levels. While we were able to build the benchmarks and install them on emulators running API Levels 23 thru 27, there is no evidence to suggest that the captured vulnerabilities can be reproduced and exploited on API levels 23 thru 27. 
Therefore, evaluations based on \droidbench should be aware of this limitation. \textit{A prudent tool evaluation strategy is to equally consider both  \droidbench and \ghera}.



\subsection{\droidbench vs \iccbench}
\paragraph*{Observtion 6} \iccbench uses only 3 APIs that are not used by \droidbench. \textit{Since \droidbench uses almost all the APIs used by \iccbench, \droidbench should be preferred over \iccbench}.

\paragraph*{Observtion 7} \iccbench benchmarks were designed to run on API level 25 whereas \droidbench benchmarks were designed to run on API levels 22 and less. Consequently, two of the three APIs related to runtime checking of permissions are used by \iccbench by not by \droidbench as these APIs were introduced in the API level 23. \textit{Therefore, if more current aspects of Android need to be considered, then \iccbench should be used in conjunction with \ghera}.


\subsection{\droidbench vs \ubcbench}
\paragraph*{Observation 8} As per \Fref{tab:api_compare_relevant_info}, \ubcbench and \droidbench share 89 APIs. \textit{Since only 10 APIs are unique to \ubcbench and \droidbench uses almost all the APIs in \ubcbench, \droidbench should be preferred over \ubcbench}.

\paragraph*{Observation 9} Nine of the 10 APIs unique to \ubcbench are related to general Java features and one API is related to SharedPreferences, a storage related feature in Android apps. This API is used by 90\% of the real-world apps and discussed by close to 1000 Android security related Stack Overflow posts which makes this API highly relevant. So, \textit{\ubcbench should be considered in conjunction with \droidbench for tools that target vulnerabilities stemming from the use of \textit{SharedPreferences}}.

\subsection{\ghera vs \iccbench}
\paragraph*{Observation 10} \ghera covers most of the ICC related APIs used by \iccbench as seen by the fact that \ghera uses 42 of the 70 APIs used by \iccbench. Furthermore, \ghera uses 476 APIs not used by \iccbench. \textit{Since \ghera uses more than 50\% of the APIs in \iccbench and is not limited to ICC, \ghera should be preferred over \iccbench}.

\paragraph*{Observation 11} \iccbench uses 28 APIs not used by \ghera. All 28 are relevant and 25 of these APIs are used by more than 60\% of real-world apps. Moreover, 23 of these APIs are related to ICC which is to be expected since \iccbench focuses on ICC. \textit{Since \iccbench uses a non-trivial number of ICC-related relevant APIs not used by \ghera, \iccbench should be considered in conjunction with \ghera}.

\subsection{\ghera vs \ubcbench}
\paragraph*{Observation 12} As per \Fref{tab:api_compare_relevant_info}, \ubcbench and \ghera share 85 APIs and \ubcbench has only 14 unique APIs. \textit{Since \ghera uses almost all the APIs used by \ubcbench, \ghera should be preferred over \ubcbench. }.

\paragraph*{Observation 13} 6 of the 14 APIs unique to \ubcbench are related to SharedPreferences and ICC while the remaining 8 are related to general Java features. \textit{Since all 14 APIs are used by more than 2000 real-world apps and 10 of the 14 APIs are used by 60\% of the real-world apps or more, \ubcbench should be considered in conjunction with \ghera when evaluating tools that target vulnerabilities stemming from the use of \textit{SharedPreferences}}.

\paragraph*{Observation 14} Similar to \droidbench, the benchmarks in \ubcbench were designed to run on API Level 19. Therefore, the authenticity of these benchmarks on API levels 23 thru 27 is unknown. Consequently, since \droidbench uses almost all of the APIs used by \ubcbench, \textit{the combination of \ghera and \droidbench should be preferred over the combination of \ghera and \ubcbench.}

\section{Answering RQ4}
\label{sec:rq4}
\textit{Do real-world apps use security-related APIs not used by any benchmark suite?}

As explained in Section \ref{sec:gapExp}, we identified 26K APIs were used by apps in our real-world app sample but not used by any benchmark.  For each of these APIs, we determined the number of Android security related Stack Overflow posts that discussed the API. We discovered that 18K of the 26K APIs are not discussed in any Android security related posts. \Fref{tab:5num-summ-real-posts} shows a five-number summary of the 8K APIs that are discussed by at least one Android security related Stack Overflow post. The numbers suggest that approximately 2K APIs (25\%) are discussed by at least 52 posts. Moreover, approximately 300 APIs are discussed in more than 1000 posts.

When we consider the package-prefixes of the 8K APIs deemed as security-related, there were 31 unique package-prefixes. From the perspective of package-prefixes, the benchmarks use APIs with only 19 of these package-prefixes; they do not use APIs with 12 of the 31 package-prefixes. We refer to these 19 package-prefixes as the \textit{known package-prefixes} and the 12 package-prefixes as the \textit{unknown package-prefixes}. \Fref{tab:covered-packages} and \Fref{tab:not-covered-packages} show that the number of APIs with \textit{known package-prefixes} is more than the number of APIs with \textit{unknown package-prefixes}. \textit{Since 60\% of the package-prefixes are known, the benchmarks are doing a good job of covering unique package-prefixes, \ie posses breadth.} However, \textit{the benchmarks are not using all security-related APIs in a known package-prefix, \ie lack depth.}

\begin{table}
  \ifdef{\TopCaption}{
    \caption{Five-Number summary of 318K security-related posts discussing APIs used in real-world apps but not in any benchmark suite. The number in parenthesis denote the absolute number of posts.}
  }{}
  \centering
  \begin{tabular}{ccccc}
  \hline
    Min & 1Q & Median & 3Q & Max\\
    \hline
    0.0002 (1) & 0.0005 (2) & 0.002 (9) & 0.13 (52) & 18.1 (69800)\\
    \hline
  \end{tabular}
  \ifundef{\TopCaption}{
    \caption{Five-Number summary of 318K security-related posts discussing APIs used in real-world apps but not in any benchmark suite.}
  }{}
  \label{tab:5num-summ-real-posts}
\end{table}

\begin{table}
  \centering
  \ifdef{\TopCaption}{
    \caption{Top 5 known package-prefixes ranked as per the number of APIs with a package-prefix}
  }{}
  \begin{tabular}{lr}
    \hline
    Package-prefix & \# APIs\\
    \hline
    android.app & 1041\\
    android.media & 772\\
    android.content & 741\\
    android.net & 530\\
    android.os & 436\\
    \hline
  \end{tabular}
  \ifundef{\TopCaption}{
    \caption{Top 5 known package-prefixes ranked as per the number of APIs  with a package-prefix}
  }{}
  \label{tab:covered-packages}

  \vspace{3mm}
  \ifdef{\TopCaption}{
    \caption{Top 5 unknown package-prefixes ranked as per the number of APIs with a package-prefix}
  }{}
  \begin{tabular}{lr}
    \hline
    Package-prefix & \# APIs\\
    \hline
    android.preference & 197\\
    android.renderscript & 141\\
    android.nfc & 133\\
    android.service & 92\\
    android.speech & 83\\
    \hline
  \end{tabular}
  \ifundef{\TopCaption}{
    \caption{Top 5 unknown package-prefixes ranked as per the number of APIs with a package-prefix}
  }{}
  \label{tab:not-covered-packages}
\end{table}

\subsection{Suggestions to extend the benchmark suites}
We wanted to know how many of the 8K APIs could be used to create new benchmarks. Consequently, we categorized the APIs based on their package-prefix to create 27 sets. We ignored 4 package-prefixes because they were related to UI and printing services that are not the focus of the benchmarks we are evaluating. Finally, we selected the top 10 APIs from each of the 27 sets based on the number of posts discussing an API to create a set of 270 APIs. We manually examined these APIs and discovered that 17 (6\%) of these APIs can be used to create new benchmarks. These APIs are related to Android app features such as ICC, loading web content, communicating via Bluetooth, interacting with a database, crypto, account management, and low-level file management.

Following are two examples of these 17 APIs can lead to vulnerabilities and can be used to create new benchmarks.

\begin{itemize}
  \item 
    Prior effort has shown that misuse of \textit{android.webkit.WebView.loadUrl(url)} can lead to vulnerability \cite{Mitra:PROMISE17}. In this exercise, we discovered \textit{android.webkit.WebView.loadUrl(url, headers)} which is a variant of \textit{loadUrl(url)} with an additional input parameter.  Given the functional similarities of these methods, \textit{loadUrl(url, headers)} can lead to similar vulnerabilities as \textit{loadUrl(url)}.

    \item The \textit{android.system.Os.open(file, mode)} API is used to access/create a file in a particular mode. If an app X saves sensitive information in a file using this API in a mode that allows other apps to access the file, then a malicious app can access this file and steal sensitive information.
\end{itemize}

Considering we found 17 APIs out of 270 manually examined APIs to lead to vulnerabilities, \textit{more APIs are likely to exist in the set of unexamined APIs could lead to vulnerabilities and be used to create new benchmarks}.

\subsection{What about APIs with no discussions?}
\label{sec:whyNoDiscs}
Out of curiosity, we explored the 18K APIs that did not appear in Android security related posts in Stack Overflow. Out of the 18K APIs, 4K were methods and the rest were fields. We examined a sample of these 4K methods.

We discovered that the method, \textit{KeyGenParameterSpec.Builder.setAttestationChallenge(bytes)}, belonging to the package \textit{android.security.keystore} is related to security. This method takes a collection of bytes as input and uses it to create an attestation challenge for a public key. An entity receiving the public key with the attestation challenge can use the challenge to verify if the public key was created in response to a specific request. However, if \textit{bytes}, needed to create the attestation challenge, is null, then the public key created will be signed with a self-signed certificate or a dummy signature. A Public key signed with a self-signed certificate or a dummy certificate is known to be insecure since such a certificate cannot be authenticated.

This API is being used by 17 apps in our real-world app sample. These 17 apps are related to finance and have a cumulative download of at least 100 million.  Clearly, this method/API is critical.  Hence, we plan to create a benchmark involving this method and contribute it to \ghera.

We also examined methods of \textit{DevicePolicyManager} class. These methods help apps control the security policies of a device such as disabling/enabling KeyGuard\footnote{KeyGuard controls the lock screen of a mobile device.} among other things. Not all apps can use the \textit{DevicePolicyManager}. Android requires apps to satisfy a particular constraint to be able to access the \textit{DevicePolicyManager}. So, an app using the \textit{DevicePolicyManager} must implement an \textit{exported} Broadcast Receiver protected by a permisson that is granted only to system entities (system permission). The methods in \textit{DevicePolicyManager} should only be accessed through the corresponding hooks in the Broadcast Receiver. For example, when the user changes the device's password the corresponding hook in the Broadcast Receiver will get invoked. The app can then implement its own logic to accept or reject the password.

While the Android documentation states that the Broadcast Receiver must be protected with a system permission, Android does not enforce it. Consequently, it is possible to implement the Broadcast Receiver without securing it with the required system permission. Hence, a malicious app without the required permission can use the Broadcast Receiver to control the device's security policies thus resulting in a privilege escalation attack. We plan to further explore these APIs.

The presence of security-related APIs in the set of 18K APIs suggests that not all security-related APIs are discussed by Android app developers on Stack Overflow. Therefore, \textit{Stack Overflow should not be used as a comprehensive source for identifying security issues in Android apps.}

\subsection{Discussion}
The APIs used in the benchmarks are discussed much more than the APIs that are used in real-world apps but not in any benchmark; see \emph{median} column in \Fref{tab:5num-summ-real-posts} and \Fref{tab:5num-summ-posts}. Additionally, almost all the filtered APIs across benchmark suites are discussed in at least one Stack Overflow post related to Android security; see \Fref{tab:5num-summ-sec-posts}. On the contrary, 75\% of the APIs used in real-world apps but not in any benchmark is not discussed in any Android security related post. Discussions around an API in the context of security implies that app developers not only use the API but also have less clarity about it. In that sense, \emph{Android app vulnerability benchmarks are doing a good job of using security-related APIs that concern app developers.}

\paragraph*{Caveat} As discussed in \Fref{sec:whyNoDiscs}, an API not being discussed in Android security related Stack Overflow posts could still be related to security. An API may not be discussed on Stack Overflow for various reasons such as 1) the API might not be directly used by app developers, 2) app developers are well aware of the security implication of the API and do not feel the need to discuss it, 3) app developers use the API but are not aware of its security implications, and 4) app developers discuss the API in another developer forum. If the APIs are not being deemed as security-related due to the third and the fourth reason, then the findings for RQ4 should be considered with caution as the 18K APIs used in real world apps but not in the benchmarks could be related to security.  This aspect does not affect the results for RQ1, RQ2, and RQ3 as very few APIs used by the benchmark suite are not discussed by Stack Overflow posts.

\section{Threats to Validity}
\label{sec:threats}
This evaluation is based on the methods and metrics proposed by Ranganath and Mitra \cite{Ranganath:EMSE19}.  Therefore, the threats to validity applicable to their effort applies to our effort as well.  According to Ranganath and Mitra, API usage is a weak measure of possibility of presence of vulnerabilities as it ignores the influence of richer but hard to measure aspects such as API usage context, security considerations of data, and data/control flow path connecting the various uses of API.  Consequently, the influence of such aspects on measuring representativeness needs to be verified. 

The baseline app from which the APIs to be ignored for representativeness calculation was determined can introduce bias based on how the baseline app was created.  The effect of this bias can be measured and mitigated by using a different baseline app.

We identified the set of security-related APIs from the set of relevant APIs using Stack Overflow data.  Specifically, we used the occurrence of APIs in posts to associate posts to APIs.  Since an occurrence of an API does not always imply the discussion of the API, the reported numbers of posts discussing an API may be inflated.  This threat can be addressed by using richer search techniques to associate posts to APIs.   Similarly, we used tags associated with posts to identify Android security related posts, and this identification can be inaccurate due to incorrect tagging of posts. This threat can be addressed by using information retrieval techniques to identify incorrect tagging.

\section{Related Work}
\label{sec:relwork}
While there has been considerable interest in developing solutions for detecting vulnerabilities in Android apps, very few efforts are focused on developing and measuring benchmarks used to evaluate the effectiveness of the solutions. Recently, Pauck \etal \cite{Pauck:2018} developed a framework for verifying the authenticity of benchmarks in DroidBench and IccBench and refining them. In a similar vein, Qiu \etal \cite{Qiu:2018} discovered that a few benchmarks in DroidBench and IccBench captured multiple aspects, which made their evaluation of the effectiveness of static taint analysis tools difficult. Therefore, they developed UBCBench and used it in conjunction with DroidBench and IccBench in their evaluation. In contrast, we focus on measuring the representativeness of benchmark suites along with comparing them and identifying gaps in them. Ranganath and Mitra \cite{Ranganath:EMSE19} performed a similar evaluation. However, their evaluation was limited to measuring the representativeness of Ghera benchmarks. Moreover, our approach of identifying security-related APIs differs from theirs as we used data from Stack Overflow to identify APIs while they manually identified APIs.

Other efforts have used API usage to categorize vulnerabilities affecting Android apps. For example, Gorla \etal \cite{Gorla:2014} used an app's description from app markets to infer its \textit{advertised} behavior and the APIs used by the app to determine any anomalies. Similarly, Sadeghi \etal \cite{Sadeghi:2017} measured the likelihood of a vulnerability pattern occurring in an Android app based on the app's source code patterns and API usage patterns. Distinct from such evaluations, the goal of our effort is to study benchmark suites and not Android apps or solutions related to Android app security.

Stack Overflow has been used in the past as a source for understanding security issues in Android apps.  Stevens \etal \cite{Stevens:2013} used Stack Overflow to study the relationship between the popularity of a \textit{permission} and the number of times a permission is overused in an app.  Similarly, Vasquez \etal \cite{Vasquez:2013} used Stack Overflow posts related to mobile development to understand the issues discussed by mobile app developers.  In a similar vein, we also used Stack Overflow to identify security-related APIs.  However, Stack Overflow data was used as an enabler and was not the focus of our evaluation.

\section{Evaluation Artifacts}
\label{sec:artifacts}

We used the version/bundle of DroidBench and IccBench benchmarks available under \textit{DroidBench (extended)} (MD5Sum 9a165494eec309ff49f1b72895308a13) and \textit{ICC-Bench 2.0} (MD5Sum: d479d07c94a9415868b420c1f289a0b2) sections at \url{https://github.com/FoelliX/ReproDroid}. The version of UBCBench we used is available at \url{https://github.com/LinaQiu/UBCBench}. The version of Ghera we used is available at \url{https://bitbucket.org/secure-it-i/android-app-vulnerability-benchmarks/src/Jan2019/}.

The raw and processed data from the experiment along with supporting scripts are available at \url{https://bitbucket.org/secure-it-i/evaluate-repr-droidbench-jan2019/src/ASE-2019/} for DroidBench, \url{https://bitbucket.org/secure-it-i/evaluate-repr-ghera-jan2019/src/ASE-2019/} for Ghera, \url{https://bitbucket.org/secure-it-i/evaluate-repr-iccbench-jan2019/src/ASE2019/} for IccBench, 
\url{https://bitbucket.org/secure-it-i/evaluate-repr-ubcbench-apr2019/src/ASE-2019/} for UBCBench,
\url{https://bitbucket.org/secure-it-i/prepare-benchmarks/src/ASE-2019/} for benchmark preparation, and
\url{https://bitbucket.org/secure-it-i/stackoverflow-march2019/src/ASE-2019/} for Stack Overflow.

\section{Future Work}
\label{sec:futureWork}
Given the increasing focus on support for securing Android apps, here are few ways to improve benchmark suites and, consequently, help improve Android security analysis tools.

\begin{enumerate}
  \item Explore richer aspects of real world apps such as call graphs, memory profiles, and API usage contexts to develop richer metrics to measure the representativeness of benchmarks.
  
  \item Explore other metrics for indentifying \textit{relevant} and \textit{security-related} APIs.
  
  \item Explore the APIs not covered by considered benchmark suites to develop and extend benchmark suites. 
  
\end{enumerate}


\section{Summary}
\label{sec:summary}

In this paper, we set out to understand how well do existing Android app vulnerability benchmark suites represent real world apps in terms of the manifestation of vulnerabilities.  We considered DroidBench, Ghera, IccBench, and UBCBench benchmark suites and used API usage as a metric to measure representativeness. 

We discovered that these benchmark suites are representative of real-world apps. Based on considered metrics, Droidbench was the most representative benchmark suite followed by Ghera, UBCBench, and IccBench. However, based on exploration of real-world APIs, we discovered that the benchmark suites are not comprehensive and they can be extending with new benchmarks. Finally, in the context of tool evaluations, the results suggest that DroidBench and Ghera should be considered equally but IccBench and UBCBench could be used to complement/strengthen the evaluations that use DroidBench and Ghera.

As an aside, we discovered that the tags associated with Stack Overflow posts are not good markers to identify posts that are likely related to security in Android apps.



\bibliography{references.bib}
\bibliographystyle{plain}

\end{document}